   \def\selectedoptions{final}

\documentclass[\selectedoptions]{aipproc}

   \def\selectedlayoutstyle{6x9}
\layoutstyle\selectedlayoutstyle

\SetInternalRegister\hbadness{8000} 

\newcommand\doingARLO[2][]{%
\ifx\mmref\undefined #1\else #2\fi}

\begin{document}

\title 
{Simple check of the vacuum structure\\ in full QCD lattice simulations}


\author{S. D\"urr}{
address={Paul Scherrer Institut, 5232 Villigen PSI, Switzerland},
email={stephan.duerr@psi.ch}}

\copyrightyear {2001}

\begin{abstract}
Given the increasing availability of lattice data for (unquenched) QCD with
$N_{\!f}\!=\!2$, it is worth while to check whether the generated vacuum
significantly deviates from the quenched one. I discuss a specific attempt to
do this on the basis of topological susceptibility data gained at various
sea-quark masses, since for this observable detailed predictions are available.
The upshot is that either discretization effects in dynamical simulations are
still untolerably large or the vacuum structure in 2-flavour QCD substantially
deviates from that in the theory with 3 (or 2+1) light quarks.
\end{abstract}

\date{\today}
\maketitle


\section{Introduction}

In a historical perspective, the path towards phenomenological predictions of
QCD by means of lattice techniques involves three steps:
Pure Yang-Mills theory (where there is just glueball physics), quenched QCD
(where the vacuum is the one in the SU(3) theory, but observables may involve
so-called {\sl current\/}-quarks) and full QCD (where the fermion determinant
with the dynamical {\sl sea\/}-quarks accounts for the quark loops in the
vacuum).
Today, the lattice community makes the final push towards full QCD, despite the
fact that state-of-the-art simulations are modestly announced as ``partially
quenched'' which means that the sea- and current-quark masses in the
(euclidean) generating functional
\begin{equation}
Z[\bar\eta,\eta]=\int DA\;\;e^{-S_G}\;\;\prod_{N_{\!f}}
\det(D\!\!\!\!\!\slash\!+\!m_\mathrm{sea})\;
\exp(\int\bar\eta{1\over D\!\!\!\!\!\slash\!+\!m_\mathrm{cur}}\eta)
\label{qcdgf}
\end{equation}
are (in general) unequal and in most cases significantly heavier than the
physical $u$- and $d$-quarks, so that phenomenological statements require a
twofold extrapolation.

Since the finite sea-quark mass constitutes the key ingredient in this
ultimate step (note that the determinant turns into a constant for
$m_\mathrm{sea}\!\to\!\infty$, hence (\ref{qcdgf}) reduces to the quenched
generating functional in that limit), an obvious task is to check whether these
``partially quenched'' or ``full'' QCD simulations exhibit the change in the
vacuum structure expected to occur if the fermions are ``active'' (i.e.\ if the
back-reaction of the ``dynamical'' fermions on the gauge background is taken
into account).
The prime observable used to distinguish the respective vacua is the
topological susceptibility
\begin{equation}
\chi(m_\mathrm{sea})={\langle q^2 \rangle \over V}\;,
\label{topsuscdef}
\end{equation}
with $q$ the (global) topological charge, because detailed theoretical
predictions show that $\chi$ behaves rather different in the quenched
($m_\mathrm{sea}\!\to\!\infty$) and chiral ($m_\mathrm{sea}\!\to\!0$) limits,
respectively.
Even though in the lattice-regulated theory (and with certain definitions of
the topological charge operator) $q$ may be somewhat ambiguous on the level of
a single configuration, the moment of the $q$-distribution which enters
(\ref{topsuscdef}) can be measured with controlled error-bars, and as a purely
gluonic object the resulting $\chi\!=\!\chi(m_\mathrm{sea})$ encodes nothing
but the vacuum structure of the theory.

Below, I give a quick survey of recent lattice determinations of $\chi$ at
various sea-quark masses in $N_{\!f}\!=\!2$ QCD, I discuss the available
continuum knowledge of the functional form $\chi\!=\!\chi(m_\mathrm{sea})$, and
I present a non-standard lattice determination of the quenched topological
susceptibility $\chi_\infty$ and the chiral condensate $\Sigma$ based on it.
The outcome is that either certain observables in todays phenomenological
studies with light dynamical quarks suffer from large discretization effects or
--~the more speculative view~-- that the low-energy structure of QCD with
$N_{\!f}\!=\!2$ is substantially different from that with $N_{\!f}\!=\!3$.


\section{Lattice Data}

I start with a quick survey of recent lattice data for the topological
susceptibility in QCD with 2 dynamical flavours; the selection reflects nothing
but my personal awareness.

{\bf CP-PACS:}
The CP-PACS collaboration has simulated full QCD on several grids at various
($\beta,\kappa$)-values, using an RG-improved gauge action and an
$O\!(a)$-improved fermion action with mean-field values for the associate
$c_\mathrm{SW}$ coefficients.
Below, I concentrate on the data generated on the $24^3\times48$ lattice at
$\beta\!=\!2.1$ with LW-cooling \cite{CP-PACS}.

{\bf UKQCD:}
The UKQCD collaboration has simulated full QCD on a $16^3\times32$ grid at
various ($\beta,\kappa$)-values, using the standard (Wilson) gauge action
and an $O\!(a)$-improved fermion action with the non-perturbative values for
the associate $c_\mathrm{SW}$ coefficients \cite{UKQCD}.

{\bf SESAM/TXL:}
The SESAM/TXL collaboration has simulated full QCD on two grids
($16^3\times32$ and $24^3\times40$) at several ($\beta,\kappa$)-values,
combining the unimproved (Wilson) gauge action with unimproved (Wilson)
fermions (i.e.\ setting $c_\mathrm{SW}\!=\!0$) \cite{SESAM}.

{\bf Thin link staggered:}
Trusting a continuum identity for the relationship between the 2- and the
4-flavour functional determinant, the staggered fermion action may be used to
simulate QCD with $N_{\!f}\!=\!2$.
There are data by the Pisa group \cite{PISA} and by A.\ Hasenfratz based on
configurations by the MILC collaboration and the Columbia/BNL project
\cite{BOULDER}.


Fig.\ 1 displays the data, along with continuum constraints to be discussed
next.

\begin{figure}
\rotatebox{90}{\resizebox{24pc}{!}{\includegraphics{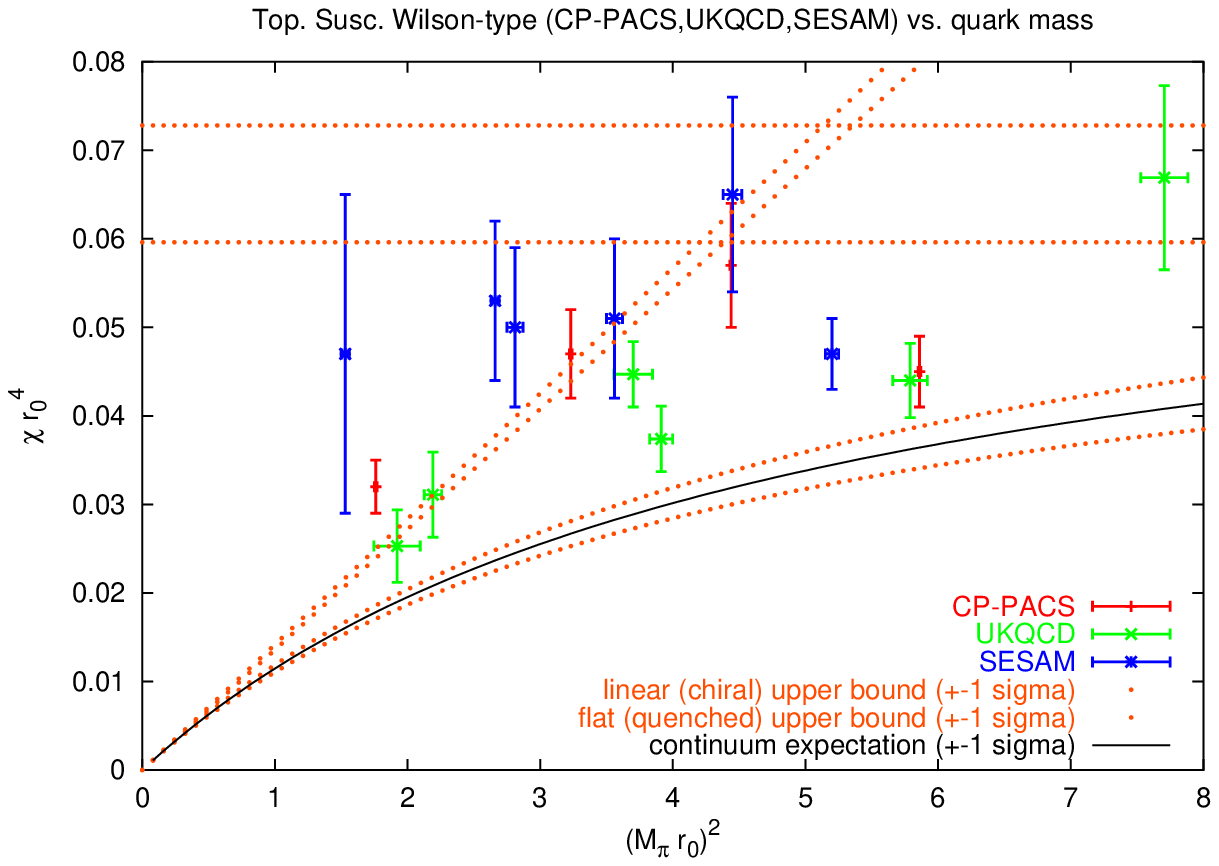}}}
\rotatebox{90}{\resizebox{24pc}{!}{\includegraphics{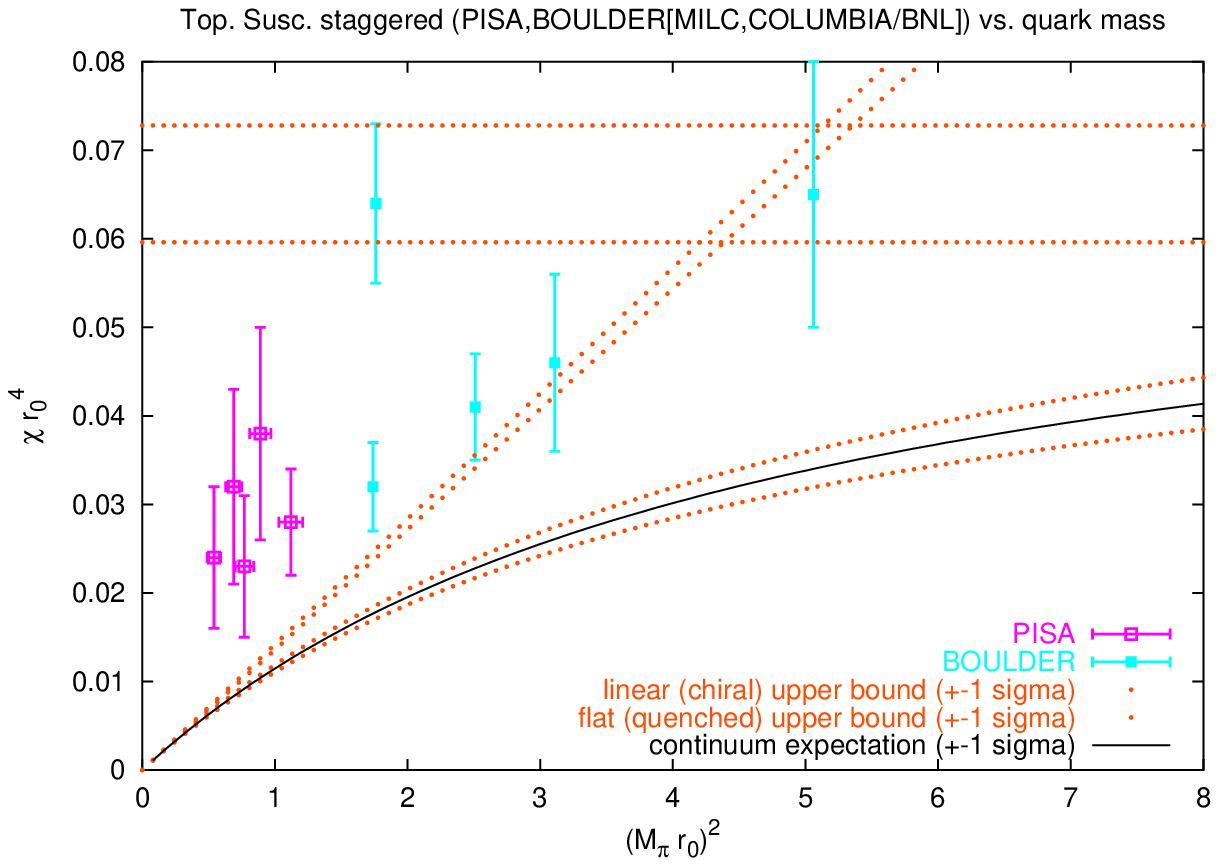}}}
\caption{Topological susceptibility versus quark mass (each in dimensionless
units) in $N_{\!f}\!=\!2$ QCD with Wilson-type (left) or staggered (right)
sea-quarks \cite{CP-PACS,UKQCD,SESAM,PISA,BOULDER}. For comparison, 1$\sigma$
bands indicating the constraints in the deep chiral regime (based on
$F_{\!\pi}\!=\!93\pm1\,\mathrm{MeV}$) and in the heavy sea-quark (quenched)
limit (from $\chi_\infty\!=\!(200\pm5\,\mathrm{MeV})^4$) are shown as well as
the associate continuum band (\ref{chiall}) (full line).}
\end{figure}


\section{Continuum Knowledge}

As mentioned in the introduction, the data for the topological susceptibility
$\chi$ versus the sea-quark mass $m\!\equiv\!m_\mathrm{sea}$ prove useful to
test the 
vacuum structure, because continuum QCD provides us with
detailed predictions:
There are analytic upper bounds for $\chi(m)$ at both asymptotically small
and large sea-quark masses and there is a ``semi-analytic'' formula for
$\chi(m)$ valid at intermediary quark masses (where the bulk of the lattice
data reside).
The only caveat is that these bounds hold true in the continuum limit, but so
far no continuum extrapolation for $\chi(m)$ in $N_{\!f}\!=\!2$ QCD is
available yet.
Before stating the complications due to this, the continuum functional forms
shall be discussed.

{\bf Asymptotically small sea-quark masses:}
For $m\!\ll\!\Lambda_\mathrm{QCD}$ and to leading order in the chiral
expansion the axial WT-identity yields (see Refs.\ cited in \cite{Durr:2001ty}
for details)
\begin{equation}
\chi(m)={\Sigma m\over N_{\!f}}\,(1\!+\!O\!(m))=
{M_{\!\pi}^2 F_{\!\pi}^2\over 2N_{\!f}}\,(1\!+\!O\!(m))\equiv
\chi_0\,(1\!+\!O\!(m))\;,
\label{chilhs}
\end{equation}
where, in the second equality, the Gell-Mann--Oakes--Renner relation has been
assumed.

{\bf Asymptotically large sea-quark masses:}
For $m\!\gg\!\Lambda_\mathrm{QCD}$ the topological susceptibility gradually
approaches its quenched counterpart (see Refs.\ cited in \cite{Durr:2001ty} for
details)
\begin{equation}
\chi(m)=\chi_\infty\,(1\!+\!O\!(1/m))=(200\pm5\,\mathrm{MeV})^4\;,
\label{chirhs}
\end{equation}

{\bf Intermediate quark masses:}
For other quark masses (i.e.\ for $(M_{\!\pi} r_0)^2\!\in\![1.5,15]$ or so,
with the Sommer scale $r_0\!=\!0.5\,\mathrm{fm}$ throughout) neither one of the
asymptotic predictions is applicable.
Fortunately, there is the ``reduced'' interpolation formula
\begin{equation}
\chi(m)=1/(1/\chi_0+1/\chi_\infty)\quad
\label{chiall}
\end{equation}
with $\chi_0$ defined in (\ref{chilhs}), which is, of course, not exact but
represents an ``educated guess''; it follows either from the chiral Lagrangian
together with pure entropy considerations (which makes it very robust)
\cite{Durr:2001ty} or from large-$N_{\!c}$ arguments \cite{Leutwyler:1992yt}.


\section{Naive evaluation of $\;\Sigma_2\;$ and $\;\chi_\infty\;$}

\begin{figure}
\rotatebox{90}{\resizebox{24pc}{!}{\includegraphics{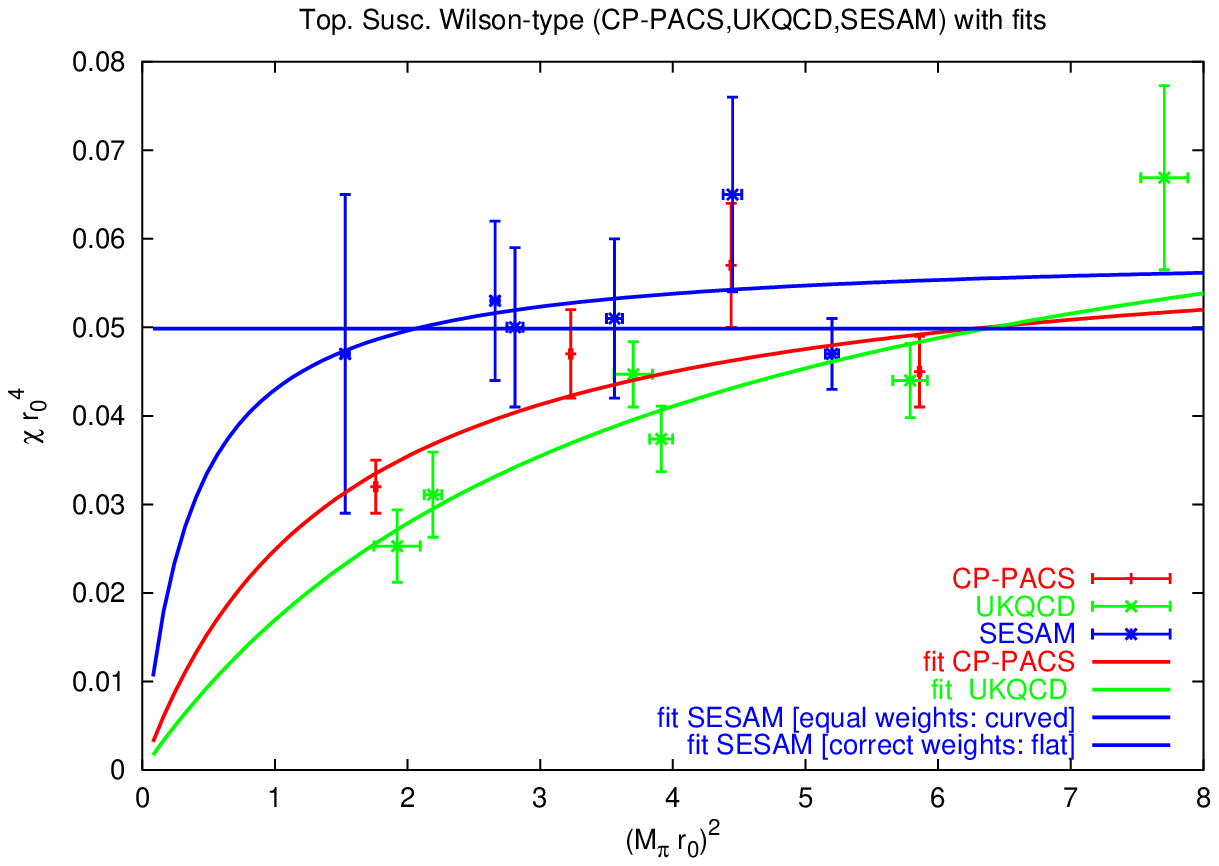}}}
\rotatebox{90}{\resizebox{24pc}{!}{\includegraphics{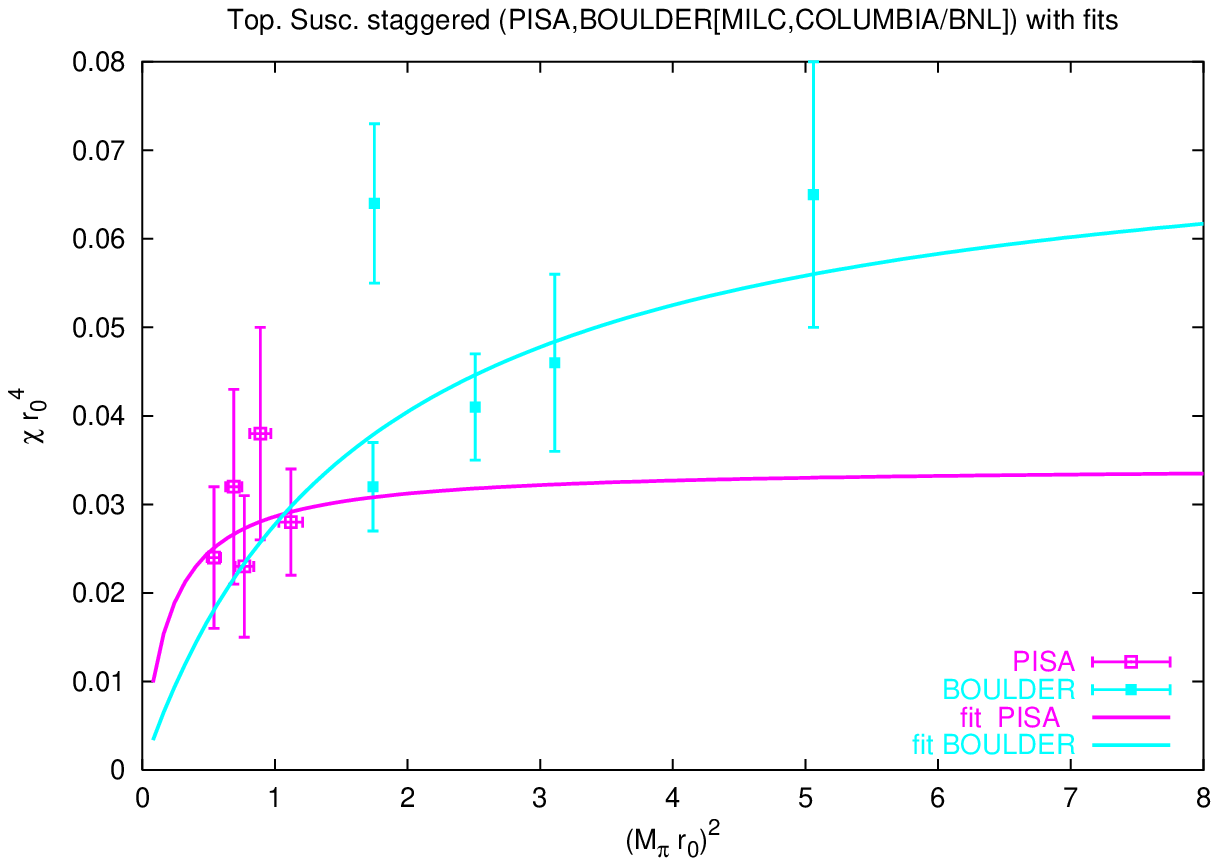}}}
\caption{Topological susceptibility versus quark mass in $N_{\!f}\!=\!2$ QCD
with Wilson-type (left) or staggered (right) sea-quarks
\cite{CP-PACS,UKQCD,SESAM,PISA,BOULDER} together with naive fits of the
susceptibility curve (\ref{chiall}) to the data, neglecting possible
discretization effects. The associate values for $\Sigma_2$ and $\chi_\infty$
are tabulated in Table~1.}
\end{figure}

Disregarding possible lattice artefacts, one may fit the available data to the
continuum curve (\ref{chiall}) and extract $\Sigma$ and $\chi_\infty$ from the
fit parameters \cite{Durr:2001ty}.
It is worth emphasizing that this evaluation of $\Sigma$ is {\sl distinct from
the usual fermionic determination\/} which is via the trace of the Green's
function of the Dirac operator at various {\sl current\/}-quark masses and
extrapolating (after proper renormalization) to the physical (or chiral) point.
Regardless how convincing this sounds, the results as tabulated in Table~1 look
rather devastating:
While our value for the quenched topological susceptibility $\chi_\infty\!
\simeq\!(200\pm10\,\mathrm{MeV})^4$ nicely agrees with previous direct
determinations in the SU(3)-theory, the suggested value for the (full) chiral
condensate in the chiral limit $\Sigma\!\simeq\!(450\pm100\,\mathrm{MeV})^3$
{\sl dramatically exceeds\/} (by more than a factor 2) the ``phenomenological''
value $\Sigma\!\simeq\!(288\,\mathrm{MeV})^3$ (which follows from the
GOR-relation with $m_{u,d}(\overline{\rm MS},\mu\!=\!2\,\mathrm{GeV})\!\simeq\!
3.5\,\mathrm{MeV}$ \cite{AliKhan:2000mw}).

Looking back at Fig.\ 1, one may argue that this hardly comes as a surprise,
since both the ``Wilson-type'' and the ``staggered'' data sets are much more
likely to violate the linear upper bound in the deep chiral regime than the
flat ceiling in the heavy-(sea)-quark limit.
Besides, Fig.\ 1 tells us how important it is to compare the data to the right
prediction: Knowing nothing but the chiral constraint (\ref{chilhs}), one might
be tempted to say that the data are in nice agreement with the leading order
chiral prediction.
The outcome of our analysis shows that there is absolutely no point in
comparing lattice data gained at $(M_{\!\pi} r_0)^2\!\geq\!1.5$ to the
leading order chiral behaviour (\ref{chilhs}), because the ``true'' prediction
(\ref{chiall}) is {\sl substantially lower\/}:
If lattice data at $(M_{\!\pi} r_0)^2\!=\!2.5$ are found to be in ``nice
agreement'' with the chiral prediction (\ref{chilhs}) based on the
phenomenological value $\Sigma\!\simeq\!(288\,\mathrm{MeV})^3$, then it means
that they are $\sim$50\% in excess of what they should be.

The bottom line is that todays full QCD simulations (with both Wilson-type and
staggered sea-quarks) --~if taken at face value~-- do show unquenching effects
in their vacuum structure but, in general, far less than expected at their
respective sea-quark masses.
Unpleasant as it is, we are invited to think about the reasons for this
finding.

\begin{table}
\begin{tabular}{|l|ccc|cc|}
\hline
{} & CP-PACS & UKQCD & SESAM \tablenote{All data get equal weights, otherwise
the best direct fit is almost flat (cf.\ Fig.\ 2).} & PISA & BOULDER\\
\hline
$(F_{\!\pi} r_0)^2/4$ & 0.0417 & 0.0216 & 0.1600 & $\;$0.1728$\;$ &  0.0441\\
$F_{\!\pi} [\mathrm{MeV}]$ \tablenote{Using the convention in which
$F_{\!\pi}\!\simeq\!93\,\mathrm{MeV}$ in nature; note that --~except for the
one in the UKQCD column~-- all entries are substantially larger than that
value.} &  161. & 116. & 316. & 328. & 166.\\
$\Sigma^{1/3} [\mathrm{MeV}]$ &  415. & 334. & 650. & 667. & 423.\\
\hline
$\chi_\infty r_0^4$ & 0.0616 & 0.0781 & 0.0587 & $\;$0.0343$\;$ & 0.0748\\
$\chi_\infty^{1/4} [\mathrm{MeV}]$ & 197. & 209. & 194. & 170. & 206.\\
\hline
\end{tabular}
\caption{Naive determination of $\Sigma$ and $\chi_\infty$ from full QCD vacuum
data with (\ref{chiall}), using $r_0\!=\!0.5\,\mathrm{fm}$ and the GOR-relation
to convert to physical units.}
\label{tabnai}
\end{table}


\section{Two alternatives}

There are two main reasons why $\Sigma$ as determined via fitting full QCD
topological susceptibility data to (\ref{chiall}) could substantially exceed
the standard phenomenological value $\Sigma\!\simeq\!(288\,\mathrm{MeV})^3$
while the simultaneously determined $\chi_\infty$ takes a regular value.

{\bf Large lattice artefacts:}
The simple reason is that discretization effects could be large, since for the
case of the topological susceptibility finite-volume effects are analytically
shown to be well under control in most of todays simulations \cite{Durr:2001ty}.
With the ascent of fermion actions which satisfy the Ginsparg-Wilson relation
it makes sense to disentangle chirality violation effects from ordinary scaling
violation effects, and in a recent paper A.\ Hasenfratz has shown some evidence
\cite{BOULDER} that one should primarily suspect the former type of
discretization effects to give rise to the excessive $\Sigma$ values listed in
Table~1.

{\bf Proximity of the ``conformal window'':}
The more ``exotic'' view is that the fitted value for $\chi_0(m)\!\simeq\!
\Sigma m/N_{\!f}$ is appropriate for $N_{\!f}\!=\!2$ QCD, while the standard
phenomenological evaluation --~even if it involves (non-strange) pions only~--
concerns QCD with 3~(2+1) light flavours.
In this scenario the way $\chi$ depends on $m$ ($m\!\ll\!\Lambda_\mathrm{QCD}$)
has an $N_{\!f}$-dependence beyond the one indicated in (\ref{chilhs}),
i.e.\ the low-energy constant $\Sigma$ depends (strongly) on $N_{\!f}$,
e.g.\ $\Sigma_2\!\simeq\!(450\,\mathrm{MeV})^3$ while 
$\Sigma_3\!\simeq\!(288\,\mathrm{MeV})^3$.
The latter gap could then be interpreted as a hint that the ``conformal
window'' (where, for appropriate $N_{\!f}$, one has $\Sigma_{N_{\!f}}\!\ll\!
\Lambda_\mathrm{QCD}^3$ and chiral symmetry is primarily broken through
higher dimensional condensates) might be ``close'' -- see \cite{Durr:2001ty}
for a discussion and some references.


\section{Refined evaluation of $\;\Sigma_2\;$ and $\;\chi_\infty\;$}

\begin{figure}
\rotatebox{90}{\resizebox{24pc}{!}{\includegraphics{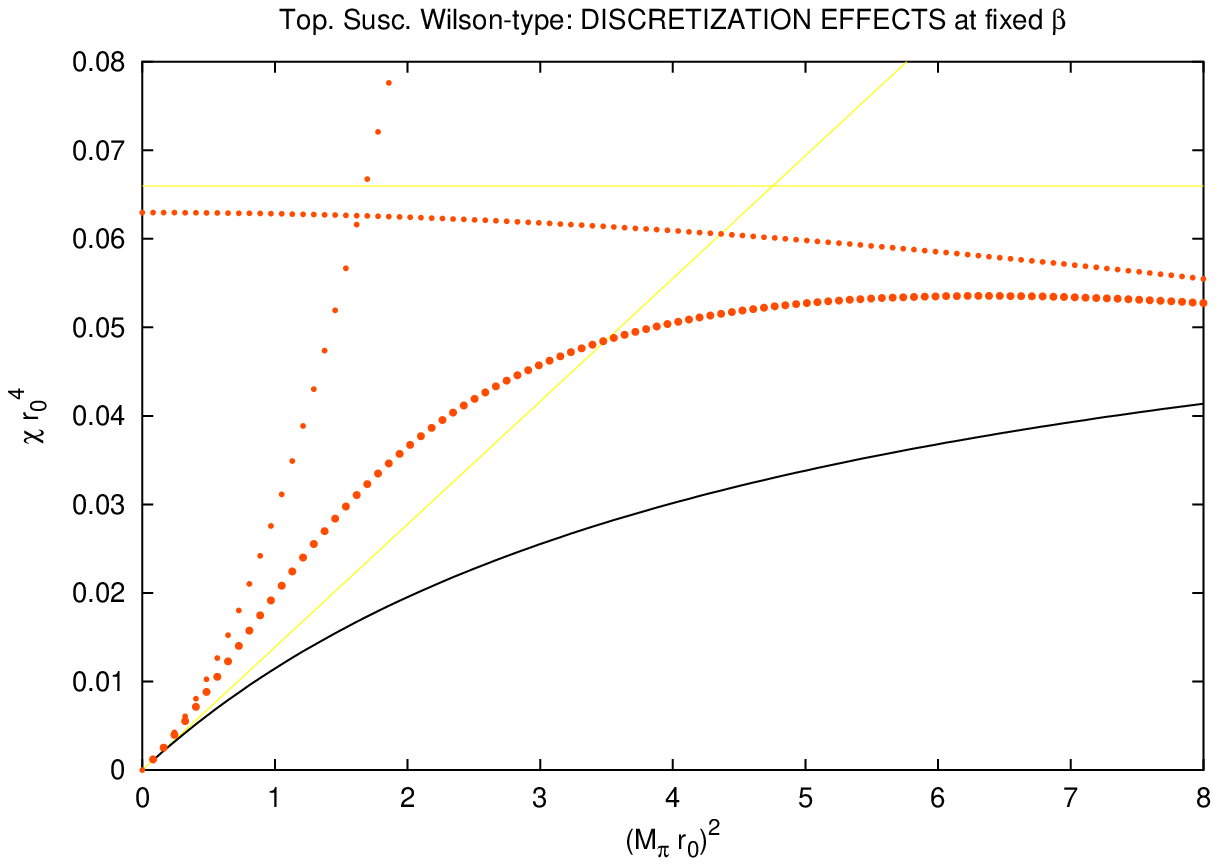}}}
\rotatebox{90}{\resizebox{24pc}{!}{\includegraphics{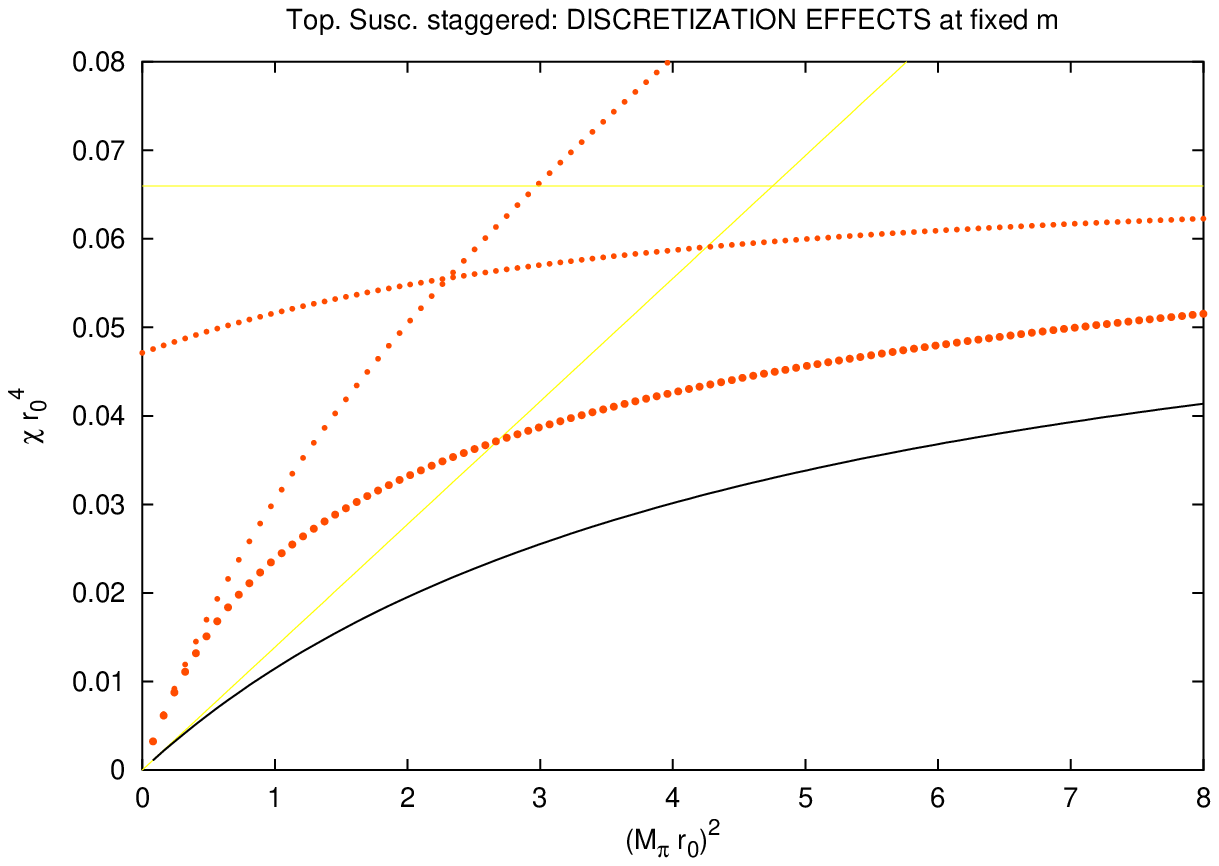}}}
\caption{Schematic representation how discretization effects typically affect
topological susceptibility data in full QCD with Wilson-type (left) and
staggered (right) sea-quarks, if simulations are run at fixed $\beta$ and fixed
$m$, respectively. In either case discretization effects tend to enhance the
effective $F_{\!\pi}$ or $\Sigma$ (in particular for large lattice spacing $a$)
and they reduce the associate $\chi_\infty$ by an amount $\propto\!a^2$. The
lattice susceptibility (\ref{chimod}) is the reduced mean of the modified
functions (fat dots) and may, for $2\!<\!(M_{\!\pi}r_0)^2\!<\!8$,
show little variation with $m$ and lie substantially above the expected
continuum curve (black line).}
\end{figure}

\begin{figure}
\rotatebox{90}{\resizebox{24pc}{!}{\includegraphics{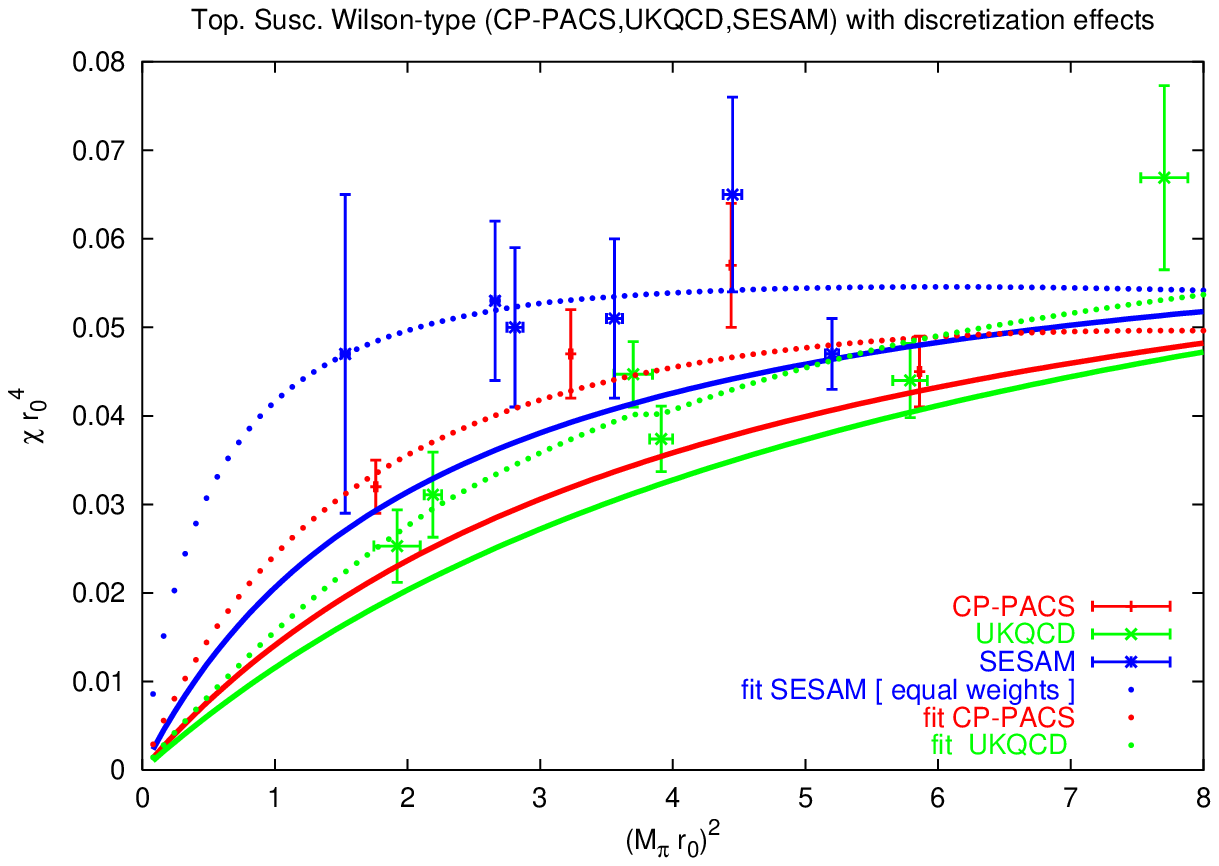}}}
\rotatebox{90}{\resizebox{24pc}{!}{\includegraphics{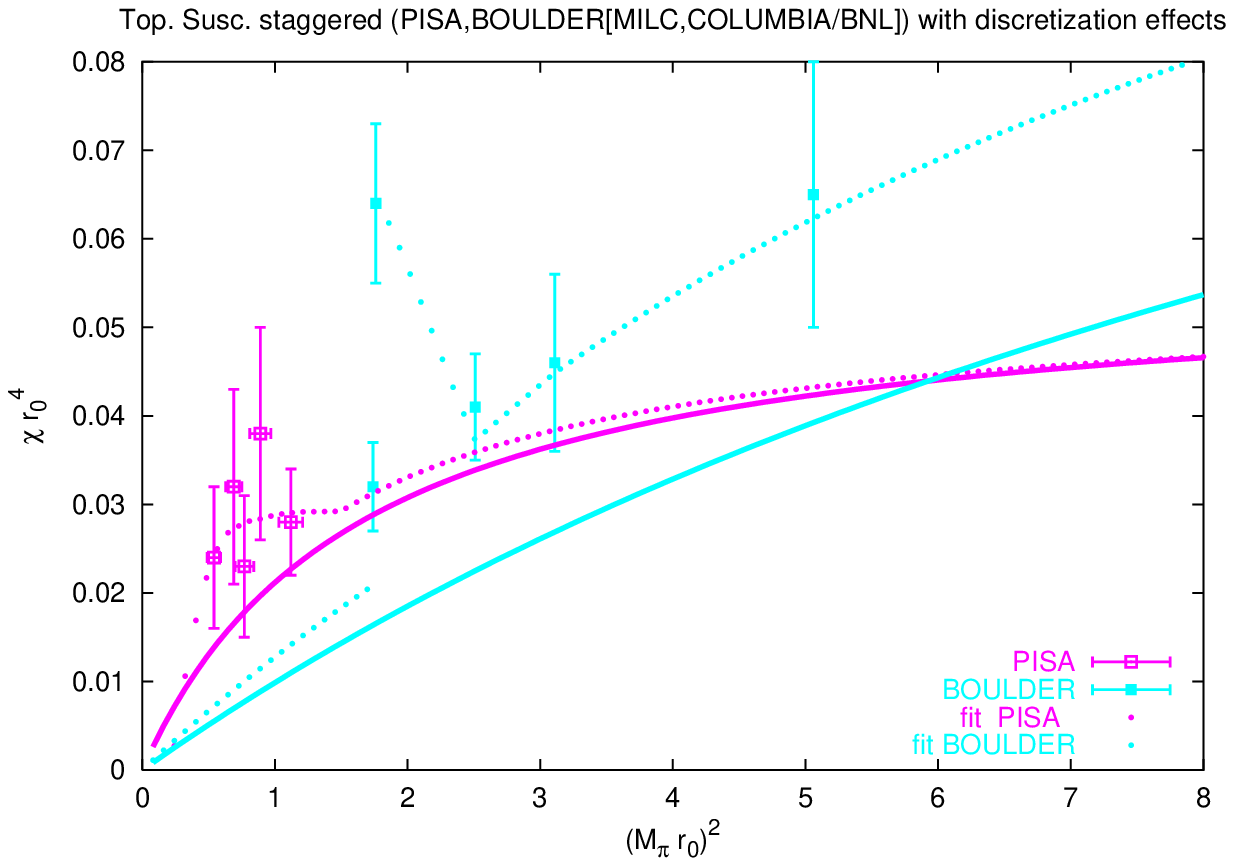}}}
\caption{Topological susceptibility versus quark mass in $N_{\!f}\!=\!2$ QCD
with Wilson-type (left) or staggered (right) sea-quarks
\cite{CP-PACS,UKQCD,SESAM,PISA,BOULDER} together with fits of the functional
form (\ref{chimod}) to the data (dotted lines). This time, the values for
$\Sigma_2$ and $\chi_\infty$ suggest ``reasonable'' continuum curves (full
lines, c.f.\ Table~2).}
\end{figure}

In the following, I concentrate on the first alternative and show that --~in
the absence of data suitable for a continuum extrapolation~-- basic knowledge
regarding the dominant discretization effects allows for a more sophisticated
evaluation of the parameters in (\ref{chiall}).

The key observation on which this analysis relies is that the leading
lattice artefacts in both elements of (\ref{chiall}) --~the chiral
piece~(\ref{chilhs}) and the quenched piece~(\ref{chirhs})~-- are known:
On the chiral side the dominant effect is chirality violation (for a discussion
see \cite{BOULDER,UKQCD}), i.e.\ $\hat F_{\!\pi}\,\hat
r_0\!=\!F_{\!\pi}\,r_0(1\!+\mathrm{const}\,(a/r_0)^{\!p})$ with $p$ reflecting
the fermion formulation.
On the quenched side scaling violations are known to result in $\hat\chi_\infty
\hat r_0^4 \!=\!\chi_\infty r_0^4 - 0.208(a/r_0)^2$ (with a known coefficient !)
\cite{UKQCD,CP-PACS}.
Combining all the ingredients, one ends up with
\begin{equation}
\hat\chi \hat r_0^4 =
1/\{\,
2N_{\!f}/[(M_{\!\pi} r_0)^2 (F_{\!\pi} r_0)^2
(1\!+\mathrm{const}\,(a/r_0)^{\!p})^q]
+
1/[ \chi_\infty r_0^4 \!-\! 0.208(a/r_0)^2 ]
\,\}\;,
\label{chimod}
\end{equation}
where $q$ may be chosen between 1 and 2, since
$(1\!+\!O\!(a^{\!p}))^2\!=\!1\!+\!O\!(a^{\!p})$; I use $q\!=\!2$.
A qualitative picture how these modifications affect the measured topological
susceptibility is drawn in Fig.\ 3.
In this respect it is important to know that in a series of full QCD
simulations at fixed $\beta$ the lattice spacing will {\sl shrink\/} if the
(sea-)quark mass gets reduced (which is often the case in studies with
Wilson-type sea-quarks; the only exception is the one by UKQCD, where $\beta$
gets relaxed when $\kappa$ is increased in such a way that $a$ stays
approximately constant), whereas if one works in a staggered formulation at
fixed quark mass~$\hat m$ the lattice will get {\sl coarser\/} as one
approaches the chiral limit -- eventually the measure resembles more that of a
theory with a single pseudo-Goldstone rather than one with $N_{\!f}^2\!-\!1$
pion type degrees of freedom as in the continuum \cite{BOULDER}.

In order to make use of this knowledge (i.e.\ to utilize (\ref{chiall}) to fit
the data) one has to decide on the parameters ($p$, const) showing up in
(\ref{chimod}).
The former choice is relatively easy -- I take $p_\mathrm{\,CP-PACS}\!=\!1.5,
p_\mathrm{\,UKQCD}\!=\!2, p_\mathrm{\,SESAM}\!=\!1,p_\mathrm{\,PISA}\!=\!
p_\mathrm{\,BOULDER}\!=\!2$ to account for the formulation and the different
strategies regarding $c_\mathrm{SW}$.
The latter choice --~which value ``const'' shall be given~-- is more delicate:
Ideally, one would like to determine it directly from the data.
However, it turns out that the quality of the data at hand is not sufficient
to allow for an additional (i.e. third) fitting parameter.
A reasonable option would be to determine it from conventional $F_{\!\pi}$
measurements on the individual ensembles.
A simpler option is to use (\ref{chimod}) twice -- in a first round ``const''
is given a likely value by fitting it while $((F_{\!\pi}r_0)^2/4,
\chi_\infty r_0^4)$ is held fixed at (0.014, 0.066); in a second round the
latter get adjusted while ``const'' is kept fixed at the previously determined
value.
Obviously, with this simpler option the final outcome for $((F_{\!\pi}r_0)^2/4,
\chi_\infty r_0^4)$ reflects, to some extent, the corresponding initial values.
The simplest option is just to set ``const'' to a generic value like 1.
My person choice is to take the arithmetic average of the ``const'' values
suggested by the last two options and to use that value to fit for
$(F_{\!\pi}r_0)^2/4$ and $\chi_\infty r_0^4$.

\begin{table}
\begin{tabular}{|l|ccc|cc|}
\hline
{} & CP-PACS & UKQCD & SESAM & PISA & BOULDER \\
\hline
$(F_{\!\pi} r_0)^2/4$ & 0.0174 & 0.0134 & 0.0299 & $\;$0.0340$\;$ & 0.0106\\
$F_{\!\pi} [\mathrm{MeV}]$ &  104. & 91. & 137. & 146. & 81.\\
$\Sigma^{1/3} [\mathrm{MeV}]$ &  311. & 284. & 372. & 388. & 263.\\
\hline
$\chi_\infty r_0^4$ & 0.0737 & 0.0845 & 0.0661 & 0.0562 & 0.1466\\
$\chi_\infty^{1/4} [\mathrm{MeV}]$ & 206. & 213. & 200. & 192. & 244.\\
\hline
\end{tabular}
\caption{Refined determination of $\Sigma$ and $\chi_\infty$ from full QCD
vacuum data with (\ref{chimod}), using $r_0\!=\!0.5\,\mathrm{fm}$ and the
GOR-relation to convert to physical units. For a cautionary statement regarding
the fitting procedure see text.}
\label{tabimp}
\end{table}

The result of this exercise is shown in Table~2 and Fig.\ 4, where dotted lines
represent the lattice curve (\ref{chimod}) with $(F_{\!\pi}r_0)^2/4$ and
$\chi_\infty r_0^4$ adjusted such as to make it go through the data points
while full lines indicate the associate continuum curve (\ref{chiall}).
The data in Table~2 should be taken with a grain of salt since, as explained
above, there is a remnant trace of the initial values in the final fitting
parameters $(F_{\!\pi}r_0)^2/4$ and $\chi_\infty r_0^4$.
Nontheless, the result is interesting because it supports the standard view
that QCD with $N_{\!f}\!=\!2$ is not in the ``conformal window'' and that its
low-energy structure agrees with that suggested by phenomenological
investigations in ``real'' (2+1 flavour) QCD \cite{Colangelo:2001sp}, i.e.\
at least for $N_{\!f}\!=\!2$ chiral symmetry is predominantly broken through a
{\sl distinctively nonzero condensate\/} (for references to an alternative
scenario see \cite{Durr:2001ty}) and $\Sigma_2$ as suggested by Table~2 is
{\sl compatible with the value from the GOR-relation\/},
$\Sigma_{2+1}\!\simeq\!(288\,\mathrm{MeV})^3$.

From a lattice perspective it reassuring to see that through a simple ansatz
for the dominant discretization effects todays state-of-the-art simulations
(which, from a naive perspective, seemed to support an almost flat topological
susceptibility curve and hence to reproduce --~in spite of all unquenching
efforts~-- a more or less quenched vacuum structure) may, in fact, be shown to
give {\sl supportive evidence in favour of a decreased topological
susceptibility near the chiral limit\/} and hence ``bear'' the knowledge of the
difference between the quenched and the unquenched vacuum structure in them.
The ultimate goal is, of course, to make discretization effects sufficiently
small so that the expected continuum pattern of the vacuum structure gets
visible in the raw data already.


\doingARLO[\bibliographystyle{aipproc}]
          {\ifthenelse{\equal{\AIPcitestyleselect}{num}}
             {\bibliographystyle{arlonum}}
             {\bibliographystyle{arlobib}}
          }
\bibliography{procmafra}

\end{document}